\documentclass[usenatbib]{mn2e}
\usepackage{graphicx}

\newcommand{\Porb}{{P_{\rm orb}}}

\newcommand{\epsneg}{\varepsilon_-}
\newcommand{\epspos}{\varepsilon_+}

\newcommand{\bv}{{\bf v}}
\newcommand{\br}{{\bf r}}

\title[A Parametric Study of Negative Superhumps]{SPH Simulations of Negative (Nodal) Superhumps: A Parametric Study}
\author[M. A. Wood et al.]{M. A. Wood$^{1,2}$\thanks{E-mail: wood@fit.edu} 
D. M. Thomas$^1$, J. C. Simpson$^3$\\
$^{1}$Department of Physics and Space Sciences,  Florida Institute of Technology, 150 W. University Blvd., Melbourne, FL 32901, USA\\
$^2$Department of Astrophysics/IMAPP, Radboud University Nijmegen, P.O. Box 9010,
6500 GL Nijmegen, The Netherlands\\
$^3$NASA  KT-C, Kennedy Space Center, FL  32899, USA}

\begin{document}

\date{Accepted 2009 June 15. Received 2009 June 12; in original form 2009 April 2.}

\pagerange{\pageref{firstpage}--\pageref{lastpage}} \pubyear{2009}

\maketitle

\label{firstpage}

\begin{abstract}
Negative superhumps in cataclysmic variable systems result when the accretion disc is tilted with respect to the orbital plane.  The line of nodes of the tilted disc precesses slowly in the retrograde direction, resulting in a photometric signal with a period slightly less than the orbital period.  We use the method of smoothed particle hydrodynamics to simulate a series of models of differing mass ratio and effective viscosity to determine the retrograde precession period and superhump period deficit $\epsneg$ as a function of system mass ratio $q$.  We tabulate our results and present fits to both $\epsneg$ and $\epspos$ versus $q$, as well as compare the numerical results with those compiled from the literature of negative superhump observations.  One surprising is that while we find negative superhumps most clearly in simulations with an accretion stream present, we also find evidence for negative superhumps in simulations in which we shut off the mass transfer stream completely, indicating that the origin of the photometric signal is more complicated than previously believed.  
\end{abstract}


\section{Introduction}

Cataclysmic variable (CV) binaries typically contain a white dwarf primary $M_1$ accreting matter from a low-mass main-sequence secondary $M_2$ via an accretion disc \citep{warner95book,hellier01}.  The Roche-lobe filling secondary star loses mass through the L1 inner Lagrange point at a rate in the approximate range $\dot M\sim10^{-11}$ to $\sim$$10^{-7}\rm\ M_\odot\ yr^{-1}$. The observed system characteristics depend most strongly on the mass ratio and individual masses, inclination, mass-transfer rate and composition, and the strength and geometry of any magnetic fields that may be present.

Systems with accretion flows that are not substantially affected by stellar magnetic fields and with mass ratios $q=M_2/M_1$ in the range $0.03 \la q \la 0.35$ (e.g., \citealt{bbw99,mmm01}; this work) can have the outer disc extend to radii approximately between that of the 3:1 corotation radius and the 2:1 corotation resonance allowing driving by the mode-coupling mechanism first identified by \citet{lubow91a}.  Within this range, common superhump oscillations can be driven and yield photometric time-series with amplitudes of up to $\sim$0.1 magnitudes and periods slightly in excess of the orbital period. 

The observed fractional period excess of the superhump oscillations increases with increasing mass ratio \citep{pattersonea05}.  The hydrodynamics of the superhump oscillations have been simulated numerically and extensively with particle-based schemes including smoothed particle hydrodynamics \citep[SPH;][]{whitehurst88,ho90,wk91,whitehurst94,murray96,murray98,sw98,wms00,fhm06, smithea07} and recently with a 2D grid-based approach \citep{kpm08}.  First recognized as photometric variations with a period a few percent longer than the orbital period that occurred during the superoutbursts of the SU UMa subclass of dwarf novae \citep{vogt74,warner75}, common superhump oscillations have since been identified in novalike CVs \citep{pattersonea93b,retterea97,skillmanea97}, the double helium white dwarf AM CVn binaries \citep{pattersonea93a,warner95amcvn,nelemans05}, and in low-mass X-ray binaries \citep{charlesea91,mho92,oc96,hynesea06}.  

Common superhumps are interchangeably referred to as {\it positive} or {\it apsidal} superhumps in the literature, or just superhumps with no modifier. 
The term ``positive'' is empirically-based and refers simply to the sign of the positive {\it period excess} $\epspos$ defined as
$\epspos =(P_+ - \Porb)/\Porb$ where we adopt the notation $P_+$ to represent the positive superhump oscillation period.
The term apsidal is dynamically-based and refers to simulation results indicating that while the radial disc profile as a function of co-rotating azimuthal angle is non-stationary, the oscillation-averaged profile is eccentric with a line of apsides that precesses slowly in the orbital prograde direction with a period $P_{\rm prec}^+$ given by
\begin{equation}
{1\over P_{\rm prec}^+} = {1\over \Porb} - {1\over P_+}.
\end{equation}

An accretion disc which is tilted out of the orbital plane will precess in the retrograde direction, analogous to the retrograde precession of the line of nodes of the moon's orbit.  We first studied this numerically in \citet[][hereafter Paper I, and see also Montgomery 2004]{wms00}.
In an early paper discussing the eclipsing binary TV Col, \citet{bbmm85} suggested that the observed 5.2-h period was the difference between the orbital frequency $\nu_{\rm orb} = 1/\Porb$ (where $\Porb=5.5$ h) and the natural retrograde precession frequency $\nu_{\rm prec}^-$ ($P_{\rm prec}^-\approx4$ d) of a tilted disc.  In a follow-up study, \citet{bow88} suggested that the physical source of the observed periodicity was the sweeping of the accretion stream bright spot across the face of a tilted disc, a model we were able to confirm in \citet[][hereafter Paper II]{wb07} and which was also discussed in \citet{fhm06}.
These photometric variations -- with periods slightly {\it less} than $\Porb$ and a completely different physical origin than the common superhumps -- have also have come to be classified as superhumps.
While the name {\it nodal} superhumps has been proposed as a physically descriptive name for these variations (and is the one we prefer), it is most common in the literature to refer to them as {\it negative} superhumps, from the sign of the relation $\epsneg = (P_- - \Porb)/\Porb$, where $P_-$ is the observed photometric period of this phenomenon.  In this work we refer to the fractional period offset of the common superhumps as the period excess $\epspos$, and the fractional period offset of the nodal/negative superhumps as the period {\it deficit} $\epsneg$.

In Paper II we presented visualizations and ray-traced simulation light curves which confirmed that the dominant physical source of the negative superhump signal is the changing depth of the accretion stream bright spot in the potential well of the white dwarf primary.  The bright spot of an untilted disc will always be located on the outer edge of the disc, but if the disc is tilted the bright spot can impact the face of the disc.  Consider that for a disc which is tilted and fixed in the inertial frame, the bright spot will impact the edge of the disc exactly twice per orbit, and between these times will sweep across first one face of the disc and then the other.  Assuming there are no dynamically-significant magnetic fields, the accretion stream will flow in the orbital plane of the binary, and impact the disc face or disc rim roughly along the line of nodes.  Negative superhump maximum occurs when the bright spot is at minimum radius -- having fallen further in the primary star's potential well, there is more specific kinetic energy to dissipate, and the bright spot is more luminous.  The ray-traced light curves in Paper II show that negative superhumps observed from opposite sides of the disc will differ in phase by $180^\circ$.
For a tilted disc that precesses slowly in the retrograde direction with a period $P_{\rm prec}^-$, the negative superhump signal has a period slightly less than the orbital period, 
\begin{equation}
{1\over P_-} = {1\over \Porb} + {1\over P_{\rm prec}^-}.
\end{equation}

\section{Numerics and Parameters}

We simulate accretion disc dynamics using our smoothed particle hydrodynamics (SPH) code\footnote{A user-friendly version of this incorporating a graphical user interface and double-buffered graphics is available for free download \citep[{\tt www.astro.fit.edu/wood/fitdisk.html};][]{wds06}.} \citep{simpson95,sw98,wb07,dws08}. 
The work presented here was calculated in parallel with a large parametric study of common superhumps to be published separately.  

SPH is a Lagrangian fluid dynamics technique originally developed by \citet{lucy77} and \citet{gm77} that has seen wide use in astrophysical applications \citep{benz90,mon92,mon05,whitehouse05,wetzstein08}.  The fluid properties are calculated by using a kernel interpolation in the scattered grid defined by the particles themselves.  As is typical, we use a normalized spline kernel  \citep{ml85} that approximates a Gaussian but is identically zero beyond an interparticle distance of $2h$, where $h$ is the {\it smoothing length}.

In their general forms, the momentum and energy equations including gravitational body forces are \citep{sw98,wb07}:
	\begin{eqnarray}
	{d^2{\bf r}\over{dt^2}}&=&
		-{\nabla P\over \rho}
		+{{\bf f}_{\rm visc}}
		-{{GM_1\over {{r_1}^3}}{{\bf r}_1}} 
		-{{GM_2\over {{r_2}^3}}{{\bf r}_2}},\\ 
		{}\nonumber\\
	{{du}\over{dt}}&=&-{P\over\rho}{\nabla\cdot\bf{v}}+{\epsilon_{\rm visc}},
	\end{eqnarray}
where $\bf {f}_{\rm visc}$ is the viscous force,
$\epsilon_{\rm visc}$ is the energy generation from viscous dissipation,
$\br_{1,2}\equiv \br - \br_{M1,M2}$ are the displacements from the stellar masses $M_1$ and $M_2$, respectively, and the remaining symbols are defined as usual.

The form of the SPH momentum equation for a particle $i$ is given by
	\begin{eqnarray} 
	{d^2{\bf r}_i\over{dt^2}}
	&=& -\sum_j m_j\left({P_i\over{{\rho_i}^2}}	+{P_j\over{{\rho_j}^2}}\right)\left(1+\Pi_{ij}\right){\nabla_i{W_{ij}}}\nonumber\\
		&&-{{GM_1\over {{r_{i1}}^3}}{{\bf r}_{i1}}}
		-{{GM_2\over {{r_{i2}}^3}}{{\bf r}_{i2}}},
	\label{sphforce}\end{eqnarray}
where $m_j$ is the mass of particle $j$ and $W$ is the SPH kernel function.
We use the artificial viscosity approximation of \citet{lattanzio86} evaluated between particles $i$ and $j$,
\begin{equation}
\Pi_{ij} = \left\{ \begin{array}{ll} 
		-\alpha\mu_{ij} + \beta\mu_{ij}^2 & {\bf v}_{ij}\cdot {\bf r}_{ij} \le 0;\\
		\\
		0  & \mbox{otherwise;}
		\end{array}\right. 
\end{equation}
where
\begin{equation}
\mu_{ij} = {h{\bf v}_{ij}\cdot {\bf r}_{ij} \over c_{s,ij}(r_{ij}^2 + \eta^2)}
\end{equation}
where $\bv_{ij} = \bv_i - \bv_j$, $\br_{ij} = \br_i - \br_j$.  The  the average of the sound speeds for particles $i$ and $j$ is $c_{s,ij}$.
For the simulations we present here, we adopt the parameters $\alpha=1.5$, $\beta=0.0$ or 0.5, $\eta=0.1h$.  As is typical in artificial viscosity prescriptions, only approaching particles feel a viscous force.  
We use a gamma-law ideal
gas equation of state $P=(\gamma-1)\rho u$ and use
$\gamma=1.01$ since we do not include radiative losses.  Within this model the sound speed is $c_s = \sqrt{\gamma(\gamma-1)u}$.

In our simulations, all particles have the same smoothing length within a given simulation, varying smoothly from $h=0.0040a$ for the $q=0.75$ run to $h=0.0067a$ for the $q=0.01$ run, where $a$ is the semimajor axis of the binary and $a=1$ in system units.  The SPH particles can have their own individual time steps as short as needed to resolve the local dynamics.  The largest timestep a particle can have is $\delta t^0=\Porb/200$, and other available timesteps are $\delta t^k = \delta t^0/2^k$, where $1 < k < k_{\rm max}$ and typically $k_{\rm max}=9$.  

As discussed in \citet{sw98}, we integrate the internal energies using an action-reaction principle which is formally equivalent to the standard SPH energy equation. 
We assume that the sum of the changes  in the internal energies of all the particles over the previous time step ($\delta t_0$) is directly proportional to the change in the bolometric luminosity of the disc over that same time interval, but do not include radiative transfer explicitly.  With this assumption we can estimate a ``simulation light curve'' as the time series over timesteps $n$, 
\begin{equation}
L^n = \sum_j \left({du\over dt}\right)_j^n \delta t_0,
\label{eq:lc}
\end{equation}
and it is the frequencies present in these light curves that we discuss below.
The dominant source of increase for the particles' internal energies is viscous dissipation, but we also include positive and negative $P\,dV$ work in the summation.  

We begin our discs {\it ab initio}, injecting at a rate of typically $R_{\rm inj}=2000$ or 3000 particles per orbit until we reach our desired total of 100,000 particles in the disc.  Our code has two options for handling particle accretion and injection.  The first option, and the one we use most often, is to promptly reinject at L1 any particles that are ejected from the system or accreted onto $M_1$ or $M_2$.  The second option decouples the injection rate from the accretion rate; injection through L1 is at a constant rate, and accreted or lost particles are not promptly replaced at the L1 region.
In the simulations we present in this work, we promptly re-inject during the initial evolution, so the total particle number simply increases linearly in time until the maximum number of particles have been injected, and from that point on the particle number is is held constant.
\begin{equation}
N = \cases{R t         & $t < N_{\rm max}/R$\\ \cr
					 & \cr
           N_{\rm max} &otherwise.\\
           }
\end{equation}
Because the discs are initially forced to have a constant total number particles as a result of the prompt particle reinjection after accretion or loss, the discs settle into a dynamical equilibrium.  Discs that are not unstable to positive superhump oscillations reach stationary states in the co-rotating frame.  Those that are driven to superhump oscillations settle into a state of {\it dynamical equilibrium} where the dissipation light curve pulse shape is stationary from one superhump cycle to the next.  

To scale our models to physical units, we use the empirical mass-radius relation of \citet{pattersonea05}.  The authors note that the data are ``too sparse to yield trustworthy results'' for secondaries with masses below $0.06 M_\odot$. However, needing an analytical relation in this regime whether trustworthy or not, we use a by-eye linear fit to the few data points shown in their Fig. 12 to obtain the modified mass-radius relation that we use to obtain the periods given in Table~1 below:
\begin{equation}
R_2 = \cases{0.23(M_2/M_\odot)^{0.27}  & $M_2 < 0.06\ M_\odot$\cr
							\cr
						 0.62(M_2/M_\odot)^{0.61}  & $0.06 < M_2/M_\odot < 0.20$ \cr
						 	 \cr
						 0.92 (M_2/M_\odot)^{0.71}  & $M_2  > 0.20 M_\odot$.}
						 \label{eq:massradius}
\end{equation}

Once the disc in a given simulation reaches a state of (dynamical) equilibrium, we impose a disc tilt of $5^\circ$ on the particle positions and velocities and restart the simulation, evolving parallel sequences both with prompt replacement and with the mass injection rate set to zero.    

\section{Results}

\subsection{Pre-tilt Evolution}

\begin{figure}
\begin{center}
\includegraphics[width=240pt]{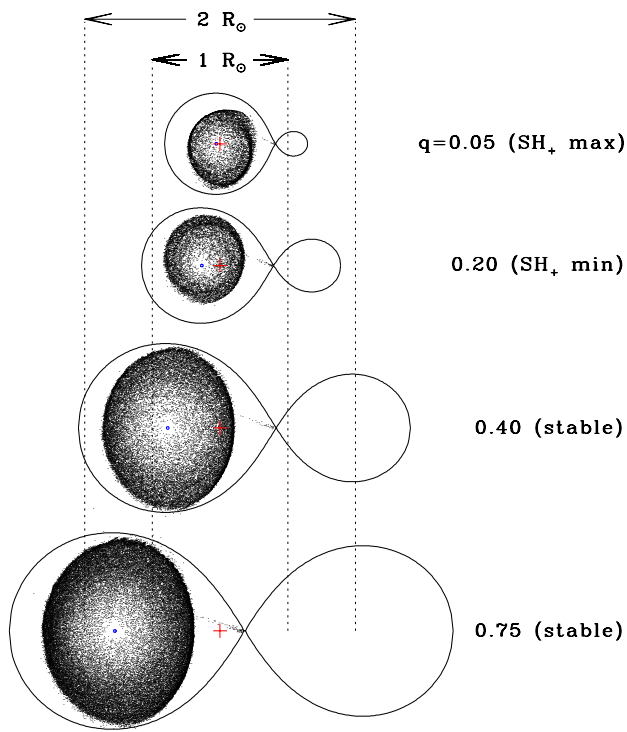}
\caption{Snapshots of mass ratios $q=0.05$ (shown near superhump maximum)  The centre of mass is indicated by a plus ($+$) sign.  The primary white dwarf is shown to scale ($R_{\rm WD}=0.011\ R_\odot$ for 0.8 $M_\odot$), and the Roche lobes are also shown.  The snapshots are at orbital phase zero.  The $x$ axis points towards the bottom of the page, the $y$ axis points to the right, and the $z$ axis points out of the page.}
\label{fig:xyq}
\end{center}
\end{figure}

We simulated systems with mass ratios from $q=0.01$ to $q=0.75$.  Simulations throughout this range yield negative superhumps in the light curves if the disc is tilted out of the plane and the simulation continues using prompt replacement of accreted/lost particles (see, e.g., Papers I \& II).  In Fig.~\ref{fig:xyq} we show ``snapshots'' of systems of 4 representative mass ratios at selected orbits and scaled to physical dimensions using the modified secondary mass-radius relation of \citet{pattersonea05} as given in Eq.~(10) above.
Our range of mass ratios includes systems with $0.03\la q\la0.35$ that are unstable to positive superhump oscillations.  In the Figure we show a disc of $q=0.05$ that is near superhump maximum, a disc of $q=0.20$ that is near superhump minimum, as well as discs of $q=0.40$ and $0.75$ that are stable.  The dimensionless disk thickness for these simulations is roughly $H/r \approx 0.04$ as determined by-eye using a plot of $z$ versus $r$.  For comparison, see Fig. 11 of \cite{wood.dqher} which shows $z$ vs.\ $r$ for a simulation of $q=0.66$ appropriate for that study of DQ Herculis.

\begin{figure}
\begin{center}
\includegraphics[width=240pt]{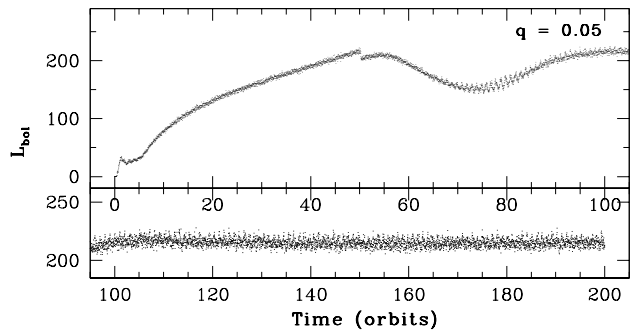}
\caption{The $q=0.05$ simulation results from start to orbit 200, by which time the disc is in a state of dynamical equilibrium defined as having common superhump oscillations that produce a light curve that repeats from one cycle to the next. In the figure above we sum 5 points in the light curve to yield 40 points per orbit, but no smoothing has been applied. Note that the scale of the lower panel differs from that of the upper panel by a factor of 2.}
\label{fig:en0.200.05}
\end{center}
\end{figure}

\begin{figure}
\begin{center}
\includegraphics[width=240pt]{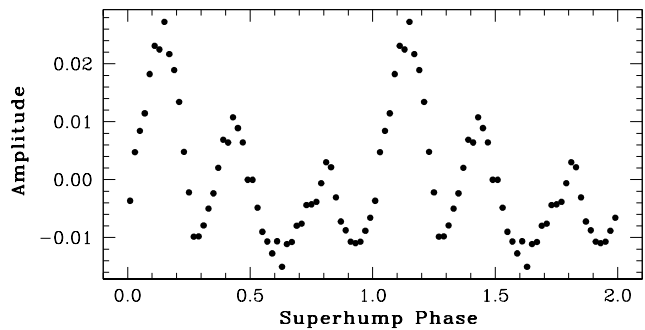}
\caption{The average apsidal superhump pulse shape for orbits 100 to 200 of the $q=0.05$ simulation.  We fold the simulation light curve on the fundamental period obtained using the Fourier transform. We bin into 50 samples per cycle, and show 2 cycles of the oscillation.  Note that the third harmonic is a dominant component to the pulse shape.}
\label{fig:en05.alc}
\end{center}
\end{figure}

\begin{figure}
\begin{center}
\includegraphics[width=240pt]{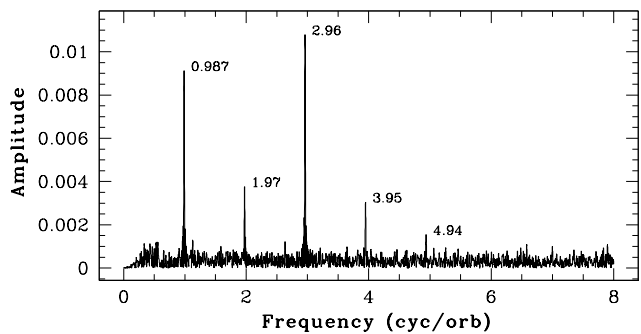}
\caption{The Fourier amplitude spectrum of orbits 100 to 200 of the $q=0.05$ simulation. The peaks in the transform are labeled with their frequencies in units of cycles per orbit.  As found in some observed systems, the fundamental (first harmonic) peak is not the highest-amplitude peak in the amplitude spectrum.}
\label{fig:en05.dft}
\end{center}
\end{figure}

As an example of a simulation unstable to common superhump oscillations, we show in Fig.~\ref{fig:en0.200.05} the light curve of 200 orbits of a $q=0.05$ simulation for which the initial injection rate was 2000 particles per orbit. Common superhump oscillations begin at approximately orbit 70, and the disc settles into a state of dynamical equilibrium by roughly orbit 100.  The average pulse shape and Fourier transform of orbits 100 to 200 are shown in Figs.~\ref{fig:en05.alc} and \ref{fig:en05.dft}, respectively.  The third harmonic dominates the average pulse shape, and this is reflected in the Fourier transform.  We will present the details of our apsidal superhump study in a separate publication.

\begin{figure}
\begin{center}
\includegraphics[width=240pt]{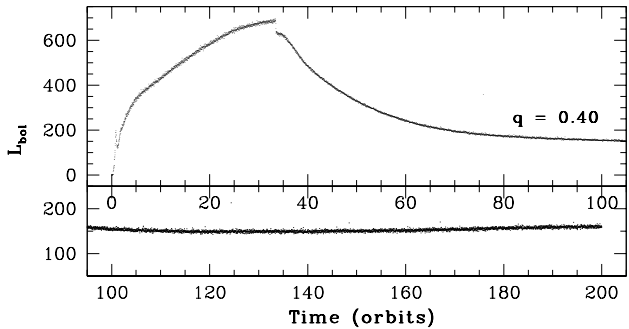}
\caption{The $q=0.40$ simulation results from start to orbit 200, by which time the disc is in a stationary state. We sum 5 points in the light curve, for a sampling of 40 per orbit. The scale of the lower panel differs from that of the upper panel by a factor of 2.}
\label{fig:en0.200.40}
\end{center}
\end{figure}

As an example of a system that is not unstable to apsidal superhump oscillations, we show in Fig.~\ref{fig:en0.200.40} the first 200 orbits of a simulation with mass ratio $q=0.40$.  The initial particle injection rate for this simulation was 3000 per orbit, so 100,000 particles have been injected by orbit 34.  By orbit 200, the disc has reached a stationary state in the co-rotating frame.  Because we note some small trend in orbits 100--200, we ran the sequence out to orbit 400 and repeated some of the experiments.  We found no difference in the resulting $\epsneg$ or simulation light curves.

The cause of the disc tilts in CVs is not completely understood at the time of this writing.  While in the low-mass X-ray binaries (LMXBs) it is believed that radiation pressure can provide the force necessary to tilt the disc \citep{petterson77,ip90,pringle96,pringle97,fhm06,ip08}, this mechanism cannot explain the disc tilt in CV discs.  \citet{bow88} suggested in their early study of TV Col that magnetic fields on the secondary might be able to deflect the accretion flow through the L1 region out of the orbital plane, thus breaking that symmetry, but as discussed in Paper II, as long as the deflection angle is stationary in the co-rotating frame and the mass transfer rate constant, the orbit-averaged angular momentum vector of the accretion stream is still parallel with the orbital angular momentum vector.  Thus no tilt could be generated. \citet{murrayea02} showed that a disc tilt could be generated by instantaneously turning on a magnetic field on the secondary star of the system.  This tilt decayed with time, but suggests that changing magnetic field geometries on the secondary could lead to disc tilts.  The model which is most generally accepted is that the primary white dwarf has a magnetic field, which can tilt the inner disc, and this information propagates outward in radius, leading to a general disc tilt.  We have completed some preliminary simulations which demonstrate that the non-parallel component (i.e., not parallel to the binary angular momentum vector) of the angular momentum  of a tilted inner disc can propagate outward in radius and result in negative superhumps. 

\subsection{Post-tilt Evolution}

\begin{figure}   
\begin{center}
\includegraphics[width=240pt]{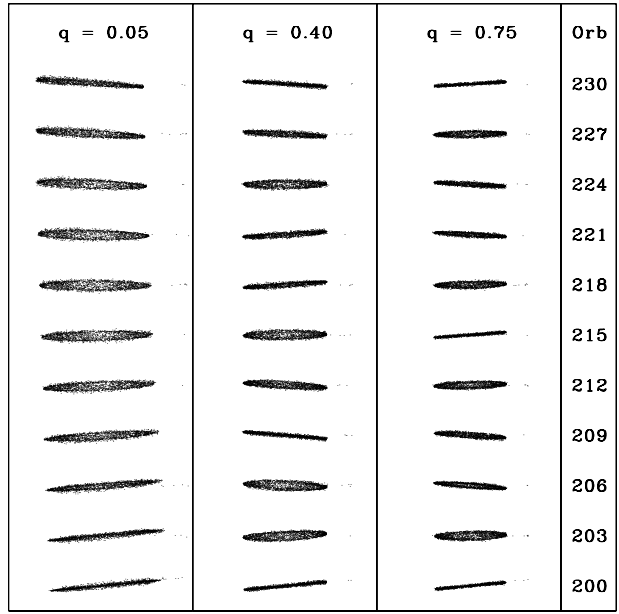}
\caption{The side-view evolution sampled every 3 orbits for simulations with mass ratios of $q=0.05$, 0.40, and 0.75, from left to right.  The Figure shows both that the nodal precession frequency is an increasing function of mass ratio and that the tilt is persistent. Note that the $q=0.05$ simulation is oscillating with common superhumps as well as precessing in the retrograde direction.   The plot limits in $x$ of each subpanel are $-0.8$ to 0.8 in system units where $a\equiv1$. The aspect ratio is 1:1 in the Figure and only 2000 randomly-selected particles are shown for each snapshot.}
\label{fig:wobble3}
\end{center}
\end{figure}

We are working on simulations that explore the origin of the disc tilt in negatively superhumping CVs, but for this work, we simply impose a $5^\circ$ tilt on a disc that has reached a state of dynamical equilibrium, and follow the evolution beyond that point.  In Fig.~\ref{fig:wobble3} we show the side-view evolution sampled every 5 orbits for sequences with mass ratios of $q=0.05$, $0.40$, and $0.75$.  The initial snapshot in each sequence shows the disc immediately after the disc tilt has been applied.  Note that while the $q=0.40$ and $0.75$ models are outside the mass ratio range that yields positive superhumps, the $q=0.05$ model shown is oscillating with positive superhumps as well as precessing in the retrograde direction.

For use in discussions below, we define the direction ``up'' to be in the direction of the orbital angular momentum vector (it is the $+z$ direction in our simulations).  This also defines the terms ``above the plane'' and ``below the plane.'' The geometry shown for the starting conditions at orbit 200 in Fig.~\ref{fig:wobble3} has the line of centres of the stars perpendicular to the line of nodes, and the disc particles nearest the secondary are above the orbital plane.  We define this to be negative superhump phase zero.   At negative superhump phase 0.5, the line of centres and line of nodes are again perpendicular, but the particles nearest the secondary are below the plane.  At negative superhump phases 0.25 (0.75) the line of centres and line of nodes are coincident, and the particles nearest the secondary have negative (positive) $z$ velocities on average.

\begin{figure}  
\begin{center}
\includegraphics[width=240pt]{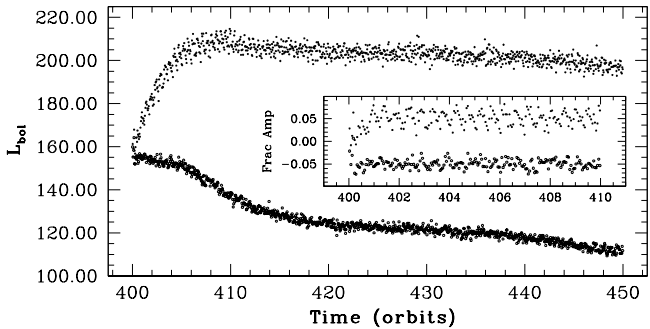}
\caption{The post-tilt light curves for the $q=0.40$ simulation.  We show both the light curve for the case of prompt replacement of accreted/lost particles (solid circles/upper curve) and for no replacement (open circles/lower curve).  Here we have summed 10 timesteps for 20 samples per orbit.  To bring out the oscillations in the light curve, the inset figure shows the first 10 orbits of the same curves as fractional amplitudes as discussed in the text. The negative superhump signal is clearly present in the upper curve of the inset, but difficult to discern in the lower curve.}
\label{fig:posttilt40}
\end{center}
\end{figure}

For each simulation run, we typically continue the evolution of the titled disc for an additional 50 orbits for the two cases of (i) prompt particle replacement and constant total number of particles in the simulation, or (ii) no particle replacement and a decreasing total number of particles in the simulation.  We show sample simulation (``bolometric'') light curves for the post-tilt evolution of these two cases for the $q=0.40$ run in Fig.~\ref{fig:posttilt40}. The upper curve of the Figure is the prompt-replacement case.   The increase in mean luminosity is the result of increased accretion onto the primary as a result of the accretion stream depositing new mass directly to annuli of small radius.  Because accreted particles are promptly replaced, this also enhances the particle injection rate at L1 over the pre-tilt average.  The system relaxes to a new quasi-equilibrium after about 10 orbits.  The lower curve of the Figure shows the case where we have shut off the injection of particles at L1 completely, which we call the ``no-stream'' case.  Here the system luminosity decreases both because there is no accretion stream bright spot, and because the total number of particles in the disc is decreasing monotonically.  At orbit 450, only 73,162 of the original 100,000 particles remain in the disc.  The overall disc geometry does not change significantly over the 50 orbits -- e.g., the root-mean-squared radius $r_{\rm rms} = \sqrt{{1\over n}\sum r_i}$ of the particle distribution is $r_{\rm rms} = 0.35a$ for both --  but because each particle has fewer nearest neighbors with which it interacts, the total viscous dissipation as a function of time is decreased.

In the inset of Fig.~\ref{fig:posttilt40} we show the fractional amplitudes of the first 10 orbits of the post-tilt evolution for the two cases.  To obtain the fractional amplitudes, we divide the original (200 samples per orbit) light curve by a copy of the light curve that has been boxcar-smoothed with a window width of 2 orbits and subtract 1.0.  The smoothed curve will contain no trace of any strictly periodic signals that have an integer number of cycles per orbit, and will essentially free of the negative and positive superhumps signals.  The residual fractional amplitude light curve will retain only frequencies higher than $\sim$$\nu_{\rm orb}$. 
Finally we then average groups of 10 points to yield the plotted light curve which effectively has 20 ``exposures'' per orbit.

In the upper light curve of the inset of  Fig.~\ref{fig:posttilt40}, the negative superhump signal can be clearly seen, while in the lower curve it is not obvious above the noise.  Note that the signal in the upper curve has a frequency of approximately 2 cycles per orbit.  As discussed in Paper II, the reason for this is that the dominant source of the negative superhump signal is the transit of the accretion stream bright spot across the face of the tilted disc.  Because our bolometric light curves reflect the viscous dissipation of {\it all} the particles in the simulation, there are negative superhump maxima corresponding to the bright spot sweeping across {\it both} faces of the disc in the prompt-replacement light curve.  The minima in this light curve correspond to the times at which the accretion stream impacts the disc rim along the line of nodes.  In an actual system, an observer only sees the effects of the bright spot sweeping across a single face of an optically-thick disc, and so sees a single maximum and single minimum per orbit.

\begin{figure}  
\begin{center}
\includegraphics[width=240pt]{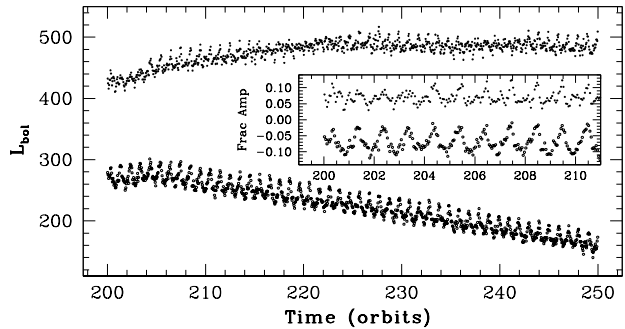}
\caption{The post-tilt light curves for the $q=0.05$ simulation.  We show both the light curve for the case of prompt replacement of accreted/lost particles (solid circles/upper curve) and for no replacement (open circles/lower curve).  Here we have summed 10 timesteps for 20 samples per orbit.  The inset figure shows the first 10 orbits of the same curves as fractional amplitudes as discussed in the text. The negative superhump signal beats with the positive superhump signal in the upper curve of the inset, but only the positive superhump signal is present in the lower curve. Positive superhumps continue long after the accretion stream is shut off, but comparison with Fig.~\ref{fig:en05.alc} shows the pulse shape has changed.}
\label{fig:posttilt05}
\end{center}
\end{figure}

For comparison, we show in Fig.~\ref{fig:posttilt05} the post-tilt evolution of the $q=0.05$ simulation, which has developed positive superhumps before the tilt is applied at orbit 200.  In this figure, we offset the no-stream light curve downward for clarity.  Note that the upper curve (with-stream) shows evidence for beating between the negative and positive superhump signals, whereas the lower curve (no-stream) shows only the positive superhump signal with no evidence for beating with a negative superhump signal.  Again, the overall luminosity declines because the total particle number and hence mean number of nearest neighbors per particle in the simulation is decreasing, which decreases the total viscous dissipation.  We note, however, that the positive superhump amplitude is nearly constant over the 50 orbits even though there is no accretion stream and hence no accretion stream bright spot.  Thus, the {\it primary} contribution to the positive superhump signal is viscous dissipation within the disc itself, and not the varying depth around the orbit of the accretion stream bright spot in the potential of the primary white dwarf as is commonly (mis)stated in the literature.  

In the inset of Fig.~\ref{fig:posttilt05}, we show the first 10 orbits of the two cases as fraction amplitudes, calculated as detailed above.  In the upper light curve, there are clearly two signals present as the light curve pulse shape evolves during the 10 orbits.  In the lower curve, only the positive superhump signal is present, and the fractional amplitude appears to increase during these 10 orbits. 
A comparison with the $q=0.05$ pre-tilt average pulse shape (Fig.~\ref{fig:en05.alc}) does show that the harmonic complexity of the light curve has decreased as the power in the third harmonic is noticeably reduced.

\subsection{Amplitude Spectra and Pulse Shapes}

A primary goal of this work is to give analytical expressions for the negative superhump period deficit $\epsneg$ as a function of mass ratio $q$ and orbital period $\Porb$.  To achieve this, we simulated systems with mass ratios spanning the range $0.01\le q \le 0.75$.  We evolved the systems to a state of equilibrium, tilted the discs by $5^\circ$ with respect to the orbital plane, and continued the simulations either by promptly replacing the accreted/lost particles, which we refer to as the ``with stream'' (ws) case, or by shutting off particle injection at L1 completely, which we refer to as the ``no stream'' (ns) case.  Simulations were continued for typically 50 orbits.  For each simulation, we take a Fourier transform of the simulation bolometric light curve converted to fractional amplitudes as described above, and after removing spurious points that are more than $6\sigma$ away from the mean.  After calculating the Fourier transform, we interpolate to find the frequencies and amplitudes of the 50 highest peaks. The peaks in the transform have a full width at half-maximum of roughly $\delta\nu\sim1/50\rm\ cyc/orb$.

\begin{figure}  
\begin{center}
\includegraphics[width=240pt]{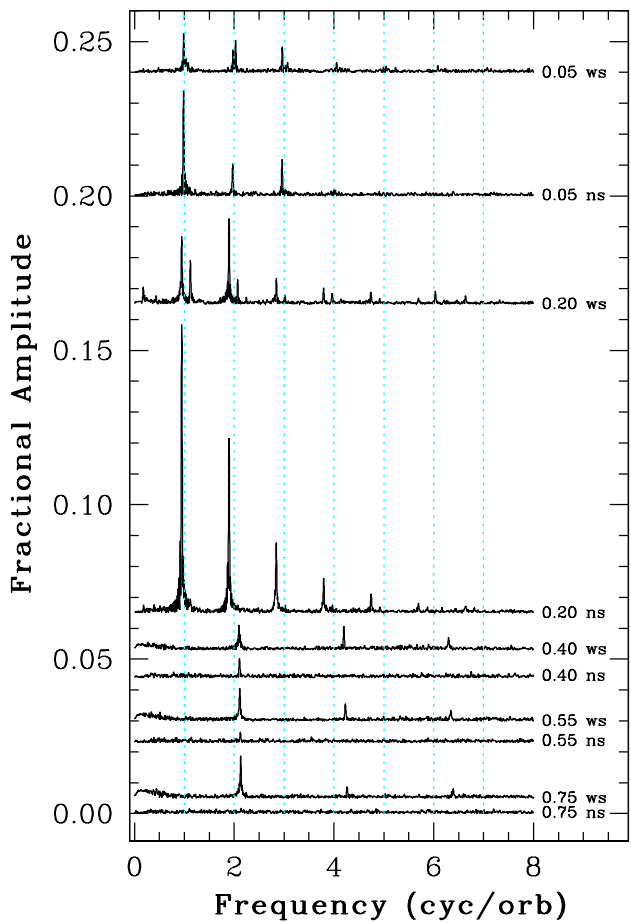}
\caption{Fourier transforms of selected post-tilt simulations.  In the figure we show Fourier transforms of 50 orbits of simulations with $q=0.05$, 0.20, 0.40, 0.55, and 0.75.  After imposing the disc tilt, we continued the simulations either with prompt replacement of accreted/lost particles (i.e., ``with stream,'' abbreviated ``ws''), or with mass injection shut off completely (i.e., ``no stream'', abbreviated ``ns''). The Fourier transforms shown are all on the same scale, but the curves have been offset for clarity.  See text for discussion.}
\label{fig:pandft}
\end{center}
\end{figure}

In Fig.~\ref{fig:pandft} we show for representative mass ratios the Fourier transforms of the post-tilt bolometric light curves in both the with-stream and no-stream cases.  The negative superhump signal in the light curves is reflected in the amplitude spectrum as a peak at a frequency slightly greater than $2\rm\ cyc\ orb^{-1}$.  This peak can be found in all spectra where the accretion stream is present, and in the spectra with mass ratios of $q=0.30$ and above in the no-stream case.  The $q=0.05$ post-tilt spectra show differences from the pre-tilt spectra shown in Fig.~\ref{fig:en05.dft} above, reflecting the differing pulse shapes.  Note that the post-tilt positive superhump first-harmonic (fundamental) amplitude is significantly higher in the no-stream case as compared with the with-stream case.  This result is even more dramatically shown in the $q=0.20$ amplitude spectra.  The positive superhump amplitudes can increase in the no-stream case because the disc is no longer accreting mass of low specific angular momentum at the disc edge, and so can expand slightly and be driven more strongly.  Careful comparison with Fig.~\ref{fig:en05.dft} shows that amplitude of the the first harmonic is also stronger in the with-stream case of the post-tilt evolution. 

In the $q=0.20$ with-stream amplitude spectrum, there is power at a frequency just above $1.0\rm\ cyc\ orb^{-1}$, but this power does not indicate a physical oscillation of the disc.  Rather, this is a combination frequency between the first harmonic of the positive superhump frequency and the negative superhump signal, resulting from non-sinusoidal pulse shapes for the two signals.  Inspection of the peak frequencies in the figure shows that the frequency difference $\Delta$ between the simulation negative superhump signal at $2\nu_-$ and the second harmonic of the positive superhump signal at $2\nu_+$ yields combination frequencies of $n\nu_+ + \Delta$ for several harmonics of the positive superhump fundamental.

\begin{figure}
\begin{center}
\includegraphics[width=240pt]{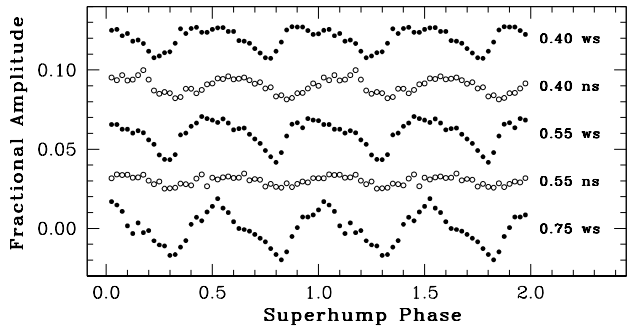}
\caption{Average pulse shapes for mass ratios $q=0.40$, 0.55, and 0.75, computed from the bolometric simulation light curves and for both the with-stream and no-stream cases (the $q=0.75$ no-stream average light curve has negligible amplitude and is not shown).  The curves show that the half-cycle pulse shapes are identical within the noise.  The character of the pulse shapes changes as a function of mass ratio, becoming more triangular for higher mass ratios for the with-stream case.}
\label{fig:aveen}
\end{center}
\end{figure}

Note that the amplitude spectra of simulations lacking positive superhumps have no significant power near $\nu = 1\rm\ cyc\ orb^{-1}$ because of the inherent symmetry discussed above.  Specifically, the transiting of the bright spot releases equal amounts of power on both faces of the disc, and the same must also be true of the tidal effects that give rise to the component of the negative superhump signal that arises even in the absence of an accretion stream. This can also be demonstrated by plotting the average pulse shapes for negative superhumps, which we show  in Fig.~\ref{fig:aveen} for a more limited but still representative sample.  To compute the average pulse shapes, we compute the fractional amplitude residuals as described above, fold the light curves on $P_-=1/\nu_-$ found from the Fourier transform, and bin into 40 bins per cycle. Because we subtract the time of the first data point from the time series before folding, zero phase in these curves is as defined geometrically above (with the line of centres perpendicular to line of nodes, and particles nearest secondary above the orbital plane).  For the mass ratios shown, it is clear that to within the noise, the simulation bolometric pulse shape repeats twice per negative superhump cycle.  Note finally that \citet{mmm09} in her study of negative superhump simulations claims that the half-cycle average pulse shapes differ (her Figs.~8 and 13).  However, her pulse shapes are undersampled and appear to contain an odd number of bins per cycle.  Her Fourier transform shows no significant power near $\nu=1$, which would have to be present for her claim to be valid.

The character of the bolometric average pulse shape changes as a function of mass ratio.  In the no-stream case, the amplitude decreases significantly from the $q=0.40$ simulation to the $q=0.55$ simulation, and is consistent with zero to the eye for the $q=0.75$ simulation (not shown).  In the with-stream case, the $q=0.40$ average pulse shape has two rounded maxima per cycle, whereas the 0.75 average pulse shape is saw-toothed.  The difference may lie in the fact that the accretion stream has less specific angular momentum for increasing $q$, and more specific kinetic energy at the bright spot (see again Fig.~\ref{fig:xyq}). 

We present the results of the Fourier analysis in Table~1, where we give for each run (defined by $q$ and viscosity coefficients $\alpha$ and $\beta$) the pre-tilt positive superhump frequency $\nu_+$ (in units of cycles per orbit) and the orbits for which that was calculated, and the post-tilt orbit range and negative and positive superhump frequencies and fractional amplitudes for both the with-stream and no-stream cases.  Entries marked with a dash (--) indicate no significant periodic signal was found. We note that the tabulated values for the negative superhump frequencies are half that found in the Fourier transforms of the bolometric light curves.  The fractional amplitudes are those of the peaks $2\nu_-$ peaks present in the transforms (see, e.g., Fig.~\ref{fig:pandft}).

\begin{table*}
 \centering
 \begin{minipage}{155mm}
  \caption{Simulation Results}
  \begin{tabular}{@{}ccccccccclccl@{}}
  \hline
  &          &    &     &\multicolumn{2}{c}{No Tilt} & {Tilt} &
    \multicolumn{3}{c}{with stream} & \multicolumn{3}{c}{no stream}\\
q & $P_{\rm orb}(h)$\footnote{Calculated using our adapted version of the empirical mass-radius relation of \citet{pattersonea05} given by Eq.~(\ref{eq:massradius}) above.}& $\alpha$ & $\beta$ &  Orbits & $\nu_+$\footnote{Frequencies are given in units of cycles per orbit $(\rm cyc/orb)$.} & 
   Orbits & $\nu_+$ &$\nu_-$ & \ \ $A_-$ &$\nu_+$ &$\nu_-$ & \ \ $A_-$ \\  
 \hline	
0.01&1.57&1.5&0.5	&	 --			&		--	&	350-400	&	  --	&	1.0047 &0.0060	&	  --	 &	-- & \ \ \ --	\\
0.03&1.44&1.5&0.5	&	550-600	&	0.990	&	600-650	&	0.987	&	1.0112 &0.0080	&	0.986 &	--	& \ \ \ --\\
0.05&1.39&1.5&0.5	&	150-200	&	0.987	&	200-250	&	0.988	&	1.0136 &0.010	&	0.985 &	--	&\ \ \  --\\
0.10&1.44&1.5&0.5	&	200-250	&	0.966	&	200-250	&	0.969	&	1.0229 &0.0084	&	0.961 &	--	& \ \ \ --\\
0.15&1.71&1.5&0.0	&	300-350	&	0.962	&	300-350	&	0.963	&	1.0289 &0.014	&	0.963 &	--	&\ \ \ --\\
0.15&1.71&1.5&0.5	&	200-340	&	0.959	&	200-350	&	0.960	&	1.0283 &0.010 	&	0.963 &	--	& \ \ \ --\\
0.20&1.94&1.5&0.5	&	400-450	&	0.946	&	400-450	&	0.948	&	1.0346 &0.0078	&	0.948 &	--	& \ \ \ --\\
0.25&2.12&1.5&0.0	&	400-600	&	0.926	&	400-450	&	0.925	&	1.0400 &0.011	&	0.924 &	--	& \ \ \ --\\
0.25&2.12&1.5&0.5 & 250-300 & 0.926 & 300-350 &	0.930	&	1.0399 &0.011	&	0.932 & --  & \ \ \ --  \\
0.30&3.27&1.5&0.0	&	400-600	&	0.903	&	400-450	&	0.903	&	1.0436 &0.011	&	0.907 & 1.0600 & 0.0041\\
0.30&3.27&1.5&0.5	&	400-500	&	0.904	&	200-250	&	0.861	&	1.0514 &0.0071	&	0.865 & 1.0471 & 0.0091\\
0.35&3.56&1.5&0.0 & 400-600 & 0.887 &	400-450	&	0.890 &	1.0460 &0.0080	&	0.900 & 1.05\footnote{Blended peak.}\ \ & 0.011\\
0.35&3.56&1.5&0.5	&		 --	  &		--	&	200-250	&	 --	  &	1.0426 &0.0063	&	0.82\footnote{Signal grows to observable amplitude at roughly orbit 240.} &	1.0488 &0.010	\\
0.40&3.82&1.5&0.5	&		--		&		--	&	400-450	&		--	&	1.0466 &0.0080	&	--	&	1.0466 &0.0062	\\
0.45&4.07&1.5&0.5	&		--		&		--	&	200-250	&		--	&	1.0497 &0.0085	&	--	&	1.0550	&0.0042\\
0.50&4.30&1.5&0.5	&		--		&		--	&	200-250	&		--	&	1.0526 &0.0097	&	--	&	1.0575	& 0.0038\\
0.55&4.52&1.5&0.5	&		--		&		--	&	200-250	&		--	&	1.0555 &0.010	&	--	&	1.0615	& 0.0033\\
0.65&4.92&1.5&0.5	&		--		&		--	&	200-250	&		--	&	1.0602 &0.012	&	--	&	1.0672	& 0.0020\\
0.75&5.28&1.5&0.5	&		--		&		--	&	200-250	&		--	&	1.0651 &0.014	&	--	&	1.0687	& 0.0017\\
\hline
\end{tabular}
\end{minipage}
\end{table*}

There are a few items of note from this table.  First, every model with steady accretion yields a negative superhump signal if the disc is tilted, and the higher the mass ratio, the higher the negative superhump frequency in cycles per orbit.   The fractional amplitudes of the negative superhumps in the with-stream case varies within a range of roughly $\sim$0.6 to 1.4\%, and those in the no-stream case vary within the range $\sim$0.17 to $1.1$\%.  The fractional amplitudes in the with-stream case do not appear to vary in a well-defined way in the range of $q$ where positive superhumps are present simultaneously, but they grow monatonically with $q$ for simulations without simultaneous superhumps, as already discussed in reference to Fig.~\ref{fig:aveen}.  

In the absence of an accretion stream, only systems with mass ratios $q\ge0.30$ display a negative superhump signal.  The fractional amplitude of the no-stream negative superhump signal rises rapidly from zero for $q\la0.25$ to $\sim$1\% for $q=0.35$, and then falls more gradually to $q=0.75$, at which point the peak in the amplitude spectrum is barely above the noise.  
We propose that the physical reason for this is 
a forced driving of the tilt instability mode initially recognized by \citet{lubow92}.  Near $q=0.30$, most of the mass in the simulated discs is near the 3:1 resonance orbit, and even at $q=0.40$, the outer disc particles complete $\sim$3 orbits per negative superhump cycle, thus periodic disipation can take place.  This driving becomes increasingly weaker as the outer disc particles move further from the resonance orbit.  Note that although for $q\sim0.35$ the {\it fractional} amplitude of the no-stream case can be larger than that of the with-stream case (which must of course also include the no-stream contribution), this is simply the result of division by the lower mean light levels for the smoothed no-stream light curves (see, e.g., Fig.~\ref{fig:posttilt40}). 

Systems in the mass-ratio range $0.03 \la q \la 0.35$ are unstable to positive superhump oscillations.  We have run a $q=0.025$ simulation out to orbit 1000 and no superhumps developed, confirming the lower bound we determined in \citet{sw98}.  Interestingly, but not surprisingly, the upper bound appears to be a function of mass ratio.  The tabulated results show that for $q=0.35$, the $\alpha=1.5$, $\beta=0.0$ simulation, positive superhumps develop, whereas they do not develop for the higher-viscosity $\alpha=1.5$, $\beta=0.5$ simulation which was run out to orbit 500.  However, once the equilibrium disc at orbit 200 is tilted, we find positive superhumps clearly developing in the no-stream post-tilt evolution at roughly orbit 240, resulting in a clear peak in the Fourier transform at $\nu_+\approx0.82$ cycles per orbit.  The with-stream result is less convincing.  There is a peak at the same location in the amplitude spectrum, but it is only about twice the height of the surrounding noise.  Finally, we note that in most cases the results for the frequencies of the negative or positive superhumps are not strongly affected by the choice of viscosity over the range we have used.

\subsection{Ray-Traced Visualizations}

\begin{figure}
\begin{center}
\includegraphics[width=240pt]{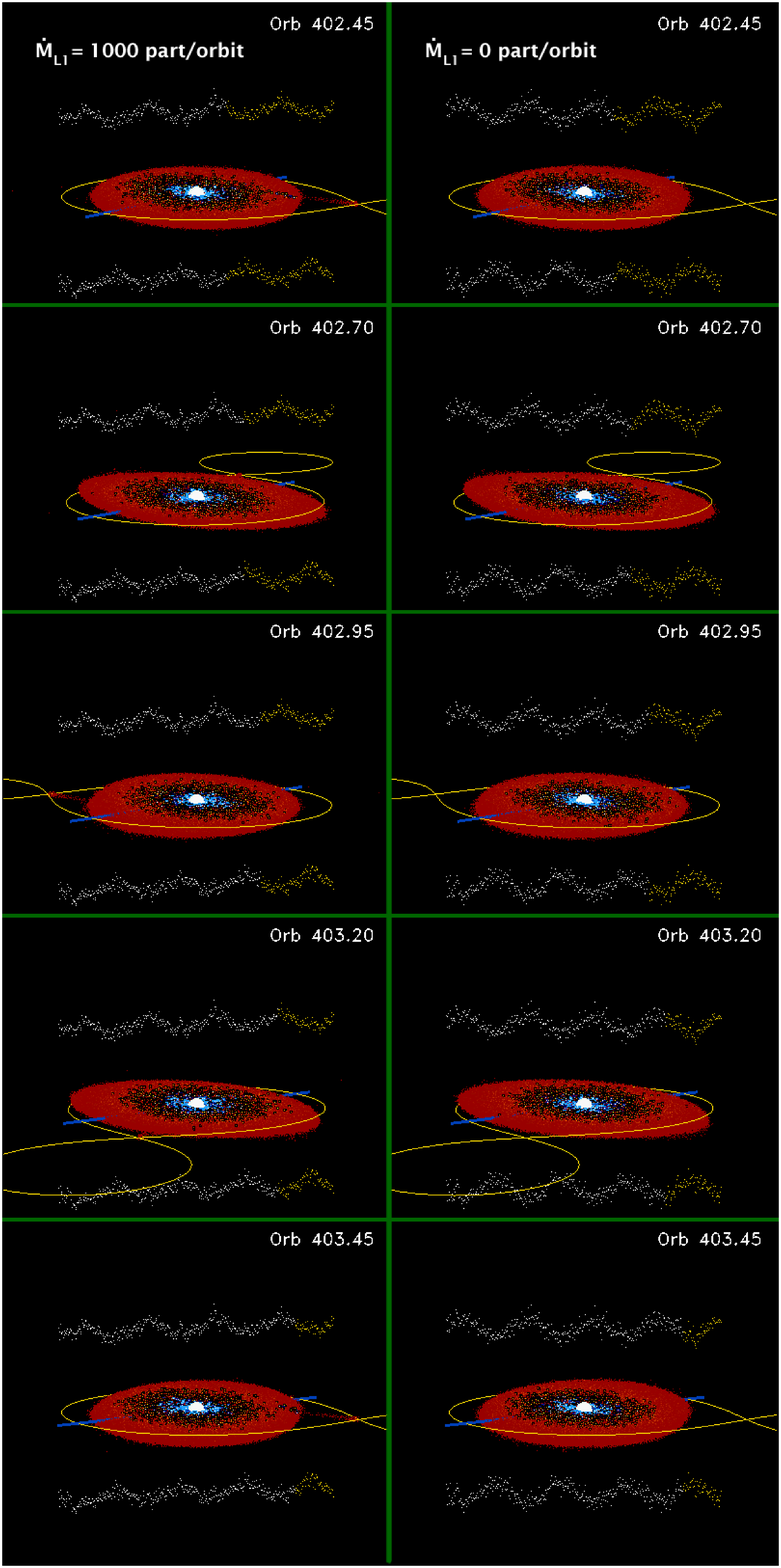}
\caption{Panel figure showing one complete orbit each of the $q=0.40$ simulations with (left panel) and without (right sequence) an accretion stream.  The system is shown in the reduced mass frame (fixed primary location).  In the Figure, the intersection of the Roche lobes with the orbital plane is shown as a yellow line, the particles are color-coded by luminosity over the previous timestep, with the brightest particles rendered as slightly-larger blue points.  The ray-traced simulation light curve calculated for $i=0^\circ$ is shown above the disc in each frame, and below for $i=180^\circ$. Note these are antiphased.  The portion of the lightcurve preceding the current timestep is a lighter color than the remainder.  The line of nodes is shown as a blue bar in the orbital plane, and it precesses in the retrograde direction.
These still frames are from animations that may be viewed as at {\tt www.astro.fit.edu/wood/visualizations.html}. 
}
\label{fig:diskpanel}
\end{center}
\end{figure}

In order to understand better the source of the negative superhump modulation, we visualized and animated our results using IDL object graphics.  We show still frames from these animations in Fig.~\ref{fig:diskpanel}.  These and other accretion disc animations can be viewed online at {\tt www.astro.fit.edu/wood/visualizations.html}.  In the figure, we include representations of the Roche-lobe profiles in the orbital plane, the primary star approximately to scale, the line of nodes, and color-code the particles by the luminosity over the previous timestep.  The light curves shown are ray-traced as described in Paper II -- we sum the energy generation (Eq.~\ref{eq:lc}) over only the ``surface'' particles along a given line of sight.  The top light curve in the figure corresponds to the $i=0^\circ$ light curve (i.e., as viewed from the positive $z$ axis), and the bottom curve corresponds to the $i=180^\circ$ light curve.  Note that these are anti-phased with respect to each other for both the with-stream and no-stream cases, and that the ray-traced light curves only show 1 maximum per superhump cycle.  The change in color of the light curves indicates the current time for that frame. 

The images in the left column of the figure result from the simulation that includes the stream, and shows 1 complete orbit.  The right column shows the same orbit for the simulation without the stream.  Note that the time value shown in the upper right hand corner of each frame is the time in orbits, not in superhump phase. The impact region of the accretion stream is visible in the with-stream images at   times 402.45 and 403.45 (top and bottom panels on the left hand side). 

\subsection{Fits to the Tabulated Results}

\begin{figure}
\begin{center}
\includegraphics[width=240pt]{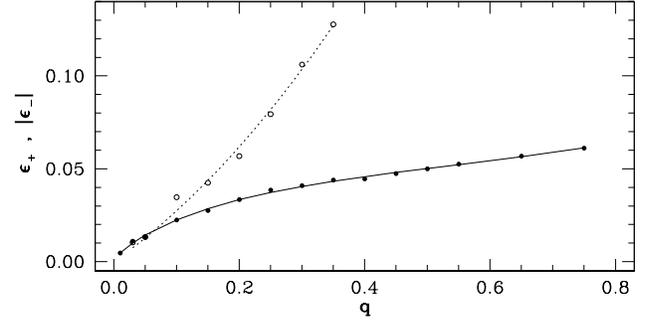}
\caption{The superhump period deficit (solid circles) and excess (open circles) as a function of mass ratio $q$.  The four-parameter fit to the negative superhump results is shown as a solid line, and the two-parameter fit to the positive superhump results is shown as a dotted line. Note that the points at $q=0.03$ and 0.05 are nearly degenerate ($\varepsilon_- \approx \varepsilon_+$) in the Table and overlap substantially in the figure.}
\label{fig:epsvq}
\end{center}
\end{figure}

In Fig.~\ref{fig:epsvq} we show the negative superhump period deficit and positive superhump period excess as functions of the mass ratio $q$. The Figure also includes fits to the numerical results.  In fitting the negative superhump results, we find that the following four-parameter fit provides an excellent approximation to the numerical results over the range $0.01 \le q \le 0.75$.  The positive superhumps are not the focus of this publication, but we include a preliminary fit to these results for completeness:

\begin{equation}
\varepsilon_- = -0.02263 q^{1/2} - 0.277 q + 0.471 q^{3/2} - 0.249 q^2
					    \label{eq:epsneg}
\end{equation}

\begin{equation}
\varepsilon_+ = 0.238 q + 0.357 q^2
	\label{eq:epspos}
\end{equation}

Because it is more common that $q$ is the desired quantity for a system with observed superhump and orbital periods, we also give fits to $q$ versus $|\epsneg|$ and $\epspos$.  We note that the apsidal precession rate depends somewhat on the disk sound speed, and that this effect becomes more important at smaller $q$.  We will explore this more fully in our forthcoming parametric study of apsidal superhumps.

\begin{equation}
q = -0.192 |\epsneg|^{1/2} + 10.37|\epsneg| - 99.83 |\epsneg|^{3/2} + 451.1 |\epsneg|^{2}
	\label{eq:epsneginv}
\end{equation}

\begin{equation}
q = 3.733 \epspos -7.898 \epspos^2
	\label{eq:epsposinv}
\end{equation}

The superhump period excess for the $q=0.10$ simulation appears to deviate substantially from the trend suggested by the remaining points, and we originally suspected that there was a problem.  However, we have checked independent simulation runs with differing values of the viscosity parameters (e.g., $\alpha=\beta=0.5$), smoothing length, and total particle number, and they consistently group near the plotted point.  We will discuss possible reasons for this in our forthcoming publication presenting the results of our positive superhump parametric study.  This is one reason why we emphasize that Eq.~\ref{eq:epspos} is preliminary.

\begin{figure}
\begin{center}
\includegraphics[width=240pt]{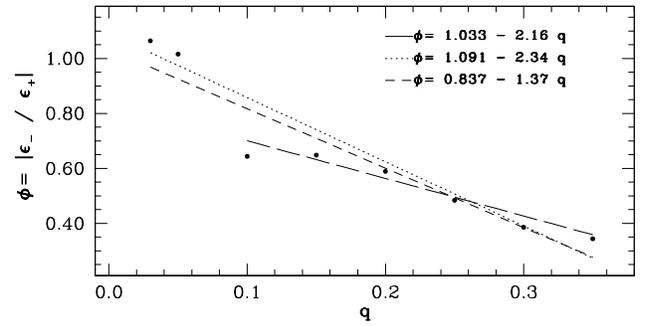}
\caption{The ratio of the superhump period deficit to excess as a function of mass ratio $q$.  Because $\epspos$ for the the $q=0.10$ simulation appears to be an outlier, three different linear fits are shown, as discussed in the text.}
\label{fig:phivq}
\end{center}
\end{figure}

In Fig.~\ref{fig:phivq} we plot the magnitude of the ratio $\phi=|\epsneg/\epspos|$ of the negative superhump period deficit to the positive superhump period excess over the mass ratio range $0.03 \le q \le 0.35$.  As expected from Fig.~\ref{fig:epsvq}, the ratio is near unity for the smallest mass ratios, and declines as mass ratio increases, a result that was hinted at in Fig.~2 of \citet{retter02}.  Again in Fig.~\ref{fig:phivq} we find that the $q=0.10$ point deviates from the trend indicated by the remaining points. Because of this we, offer 3 different linear fits to the results.  Fitting all the data points we obtain

\begin{equation}
\phi = 1.033 - 2.16 q.
\end{equation}
If we chose not to include the $q=0.10$ point, then the linear fit is
\begin{equation}
\phi = 1.091 - 2.34 q.
\end{equation}
Finally, in consideration of the fact that negative superhumps are primarily only {\it observed} in longer-period systems (i.e., above the period gap), we have a fit to the points in the range $0.10 \le q \le 0.35$,
\begin{equation}
\phi = 0.837 - 1.37 q.
\end{equation}

\begin{figure}
\begin{center}
\includegraphics[width=240pt]{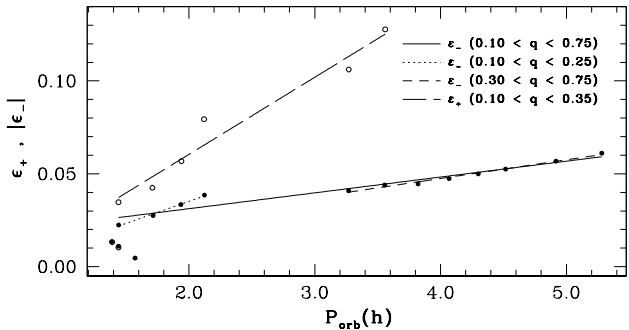}
\caption{The superhump period deficit (solid circles) and excess (open circles) versus orbital period calculated assuming $M_1=0.8 M_\odot$ and the secondary mass-radius relation given by Eq. (\ref{eq:massradius}). The linear fits are discussed in the text.}
\label{fig:evporb}
\end{center}
\end{figure}

Up to this point we have presented our results as a function of mass ratio, because that is the fundamental parameter in the simulations.  However, the mass ratio is much more difficult to measure observationally than orbital or superhump periods, and so in Fig.~\ref{fig:evporb} we show the negative (positive) superhump period deficit (excess) as a function of $\Porb$, where we used the modified \citet{pattersonea05} secondary mass-radius relation (Eq.~10) to scale our mass ratios to periods, assuming that the white dwarf primary mass is $M_1=0.8 M_\odot$.  This empirical mass-radius relation yields a minimum $\Porb$, and for this reason we confine our fits to mass ratios of $q\ge0.10$.
These relations should help guide observers to determine if a putative photometric period detection is plausibly the result the negative superhumps. 
Fitting all the negative superhump results over the range $0.10\le q\le0.75$, we obtain
\begin{equation} 
\epsneg = -0.0142   - 0.00852 \Porb.
\end{equation}
Because the negative superhump results appear nearly linear over the range $0.10\le q\le0.25$, we fit these and the remaining points separately
\begin{equation}
\epsneg = \cases{\ 0.01225 - 0.0237\Porb & $0.10 \le q \le 0.25$,\cr
 								\cr
                 -0.00722 - 0.0101\Porb & $0.30 \le q \le 0.75$.\cr
}
\end{equation}
Finally, the best linear fit to the positive superhump period excess over the range $0.10 \le q \le 0.35$ is
\begin{equation}
\epspos = -0.0226 + 0.0415 \Porb.
\end{equation}

\section{Observational Results}

\subsection{Systems with Main Sequence Secondaries}

\begin{figure}
\begin{center}
\includegraphics[width=240pt]{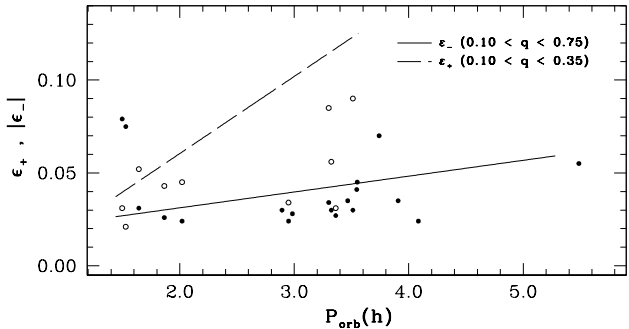}
\caption{Observed superhump period deficit (solid circles) and excess (open circles) versus orbital period for systems listed in Table 2.  The fits are as presented in Fig.~\ref{fig:evporb}.  The scatter of the observations is substantial.  We do not include the ultra-compact systems AM CVn or V1405 Aql in the figure as the secondaries in these systems are not main sequence stars, and so these systems follow a different $q$ vs.\ $\Porb$ relation.}
\label{fig:evporb-obs}
\end{center}
\end{figure}

In Table 2 we list 21 systems that have published negative superhump period deficits. For each system, we list the orbital period, the published period deficit, and the published period excess if it exists.  The data are presented graphically in Fig.~\ref{fig:evporb-obs}.  In the figure, the solid circles represent the negative superhump period deficits, and the open circles the period excesses.  While most of the reported period deficits are found near the best-fit relationship we find from our numerical results, a handful of the observed period deficits are quite discrepant.  In general the systems require very long photometric time series observations over a period of a week or more and at multiple longitudes for best results. It may be useful to revisit several of these systems to determine if the published period deficits can be confirmed.

\begin{table}
 \begin{center}
  \caption{Observational Results.}
  \begin{tabular}{@{}llllll@{}}
  \hline
Object &  $P_{\rm orb}$(h) & $\varepsilon_-$ & $\varepsilon_+$ & Refs\\  
 \hline	
AM CVn					&	0.286	&	-1.7	&	2.2	&	1, 2, 3\\
V1405 Aql				&	0.834	&	-0.71 &	0.90	&	4, 5, 6\\
V1159 Ori				&	1.492	&	-7.6	&	3.2	&	7, 8\\
ER UMa					&	1.528	&	-7.5	&	2.1	&	9, 10\\
IR GEM					&	1.642	&	-3.1	&	5.2 &	11\\
V503 Cyg				&	1.865	&	-2.6	&	4.3	&	12\\
BF Ara					&	2.020	&	-2.4	&	4.5	&	13\\
RX J1643+3402		&	2.893	&	-3.0	&	--	&	14\\
AH Men					&	2.950	&	-2.4	&	3.4	&	15, 16\\
V442 Oph				&	2.984	&	-2.8	&	--	&	14\\
TT Ari					&	3.301	&	-3.4	&	8.5	&	17, 18, 19\\
V603 Aql				&	3.314	&	-3.0	&	5.6	&	20\\
RR Cha					&	3.362	&	-2.7	&	3.1	&	21\\
V751 Cyg				&	3.467	&	-3.5	&	--	&	22, 23\\
PX And					&	3.512	&	-3.0	&	9.0	&	3, 24\\
V2574 Oph				&	3.546	&	-4.1	&	--	&	25\\
HS 1813+6122		& 3.55  & -4.5  & --  & 26\\
BH Lyn					&	3.741	&	-7.0	&	--	&	27\\
KR Aur					&	3.907	&	-3.5	&	--	&	28\\
SDSS J0407-0644	&	4.084	&	-2.4	&	--	&	29\\
TV Col					&	5.486	&	-5.5	&	15  &	30, 31\\
\hline
\end{tabular}
\end{center}
\raggedright
\noindent {\it References.} 
1 \citet{skillman99};
2 \citet{patterson98};
3 \citet{patterson99};
4 \citet{chou01};
5 \citet{hu08};
6 \citet{retter02};
7 \citet{pattersonea95};
8 \citet{thorstensen97};
9 \citet{gao99};
10 \citet{zhao06};
11 \citet{fu04};
12 \citet{harvey95};
13 \citet{olech07};
14 \citet{pattersonea02};
15 \citet{patterson95};
16 \citet{rodriguez-gil07};
17 \citet{skillmanea98};
18 \citet{andronov99};
19 \citet{wu02};
20 \citet{pattersonea97};
21 \citet{ww02};
22 \citet{pattersonea01};
23 \citet{papadaki08};
24 \citet{stanishev02};
25 \citet{kang06};
26 \citet{rodriguez-gil07b};
27 \citet{stanishev06};
28 \citet{koz07};
29 \citet{ak05b};
30 \citet{hellier93};
31 \citet{retter03};
\end{table}

Negative superhumps have been tentatively identified in a handful of additional systems, but we did not include these systems in Table 2 and Fig.~\ref{fig:evporb-obs}.
\citet{tramposch05} suggest that SDSS J210014.12+004446.0 shows negative and positive superhumps with periods of
$1.96\pm0.02$ and $2.099\pm0.002$, respectively.  The scatter in their phase-resolved spectroscopy suggests an orbital period near 2 hr, but the value is not well determined.  
\citet{ak05a} suggested that V1193 Ori displayed negative superhumps in their photometric time series observations, but to conclude that they had to assume that the actual orbital period was a 1 $\rm d^{-1}$ alias of the published value of $\Porb = 0.165$~d \citep{ringwald94,papadaki04}.  \citet{papadaki06} were able to confirm the published value of the orbital period and were not able to confirm any of the periods reported by \citet{ak05a}.  
The long-period system TX Col was claimed to potentially show both positive and negative superhumps by \citet{retter05} based on single-site data, however they warn that their observations need confirmation. 

\subsection{V1159 Ori and ER UMa}

\citet{pattersonea95} show evidence of a 82.7-min periodicity in V1159 Ori which occured on two separate nights, and they identify this as a negative superhump signal.  Using the spectroscopically-determined $\Porb = 89.5363$~min \citep{thorstensen97}, this yields a period deficit of $-7.6$\%, where we'd expect a value closer to $\epsneg\sim-1.6$\% based on our results.  
Similarly, \citet{gao99} and \citet{zhao06} report finding negative superhumps in ER UMa also with a very large period deficit of $\epsneg=-7.5$\%.  Here the period is found to vary night-to-night, and the run lengths are short.
If these signals can be confirmed, then they are perhaps the first observed signals at $\nu_+ + \Delta$ discussed above in reference to the $q=0.20$ with-stream amplitude spectrum (Fig.~\ref{fig:pandft}). However, the absence for either system of a detected signal near what should be the ``true'' $2\nu_-$ weakens this interpretation. Additional observations are clearly needed.

\subsection{AM CVn \& V1405 Aql}

AM CVn is perhaps the best studied of the helium-mass-transfer binaries, and the prototype for the class \citep{nelemans05}.  The most extensive photometric study to date for this object is that of \citet{skillman99}. The positive superhump excess is $\epspos=2.2\%$, which using Eq.~(\ref{eq:epsposinv}) above yields a mass ratio estimate of $q\approx0.08$.  \citet{skillman99} report $\epsneg=-1.7$\% for AM CVn, which from Eq.~(\ref{eq:epsneginv}) above yields an estimate of $q\approx0.06$. These estimates are consistent with each other and with previous estimates based the observed $q$ versus $\epspos$ relationship \citep[e.g.][]{pattersonea95}.  
However, more recently, \citet{roelofsea06} presented the results of an extensive time-series spectroscopic study of AM CVn in which they find a kinematic feature that appears analogous to the ``central spike'' feature observed in some longer period AM CVn stars.  If, as in these other systems, the observed central spike of the He II 4471 line traces the projected velocity and phase of the accreting white dwarf, then $q=0.18\pm0.01$ for AM CVn \citep{roelofsea06}, which is substantially higher than the $q$ inferred from the superhump signals.  \citet{pearson07} explores the possibility that helium discs in AM CVn systems may have pressure effects operating in their discs that differ significantly from those characteristic of hydrogen-dominated discs, and this coupled with possible differing secondary star formation channels for these systems \citep[see, e.g.,][]{roelofsea07,yungelson08} may preclude the existance of a well-defined superhump period excess to mass ratio relationship.  However, other AM CVn stars with measured superhump and orbital periods generally appear to be consistent with the hydrogen-disc results, so additional effort is required to resolve these issues.

V1405 Aql is a LMXB first discovered as an X-ray source exhibiting type-I bursts, suggesting a neutron-star primary \citep[see][and references therein]{retter02}.   Observations indicate an optical period about 1\% longer than the 0.834-h X-ray orbital period, consistent with common superhumps, and \citet{retter02} report the detection of a periodicity 0.828 h, which they claim to result from negative superhumps.  Thus, for V1405 Aql the negative superhump period deficit is $\epsneg=0.71\%$ which implies $q=0.021$ using Eq.~(\ref{eq:epsneginv}) above, and the positive superhump period excess is $\epspos=0.90\%$, for which Eq.~(\ref{eq:epsposinv}) implies $q=0.033$.  For a $1.4~M_\odot$ neutron star primary, these results suggest a secondary mass in the range 0.03--0.05~$M_\odot$, and because we find positive superhumps only for $q\ge0.03$, we would suggest the secondary star has a mass near $0.05~M_\odot$. Given the discrepency for AM CVn between the spectroscopically-determined mass ratio and that inferred from the superhump periods, the above determinations must of course be taken as preliminary.

\section{Conclusions}

We report on the results of an extensive parametric study of negative superhumps using the method of smoothed particle hydrodynamics.  Negative superhumps are visible when a cataclysmic variable system has a disc which is tilted out of the orbital plane.  Torques operating on the tilted accretion disc cause it to precess in the retrograde direction, at a rate that increases with system mass ratio $q$.

Our results reveal that the negative superhump signal is dominated by the transit of the accretion stream impact spot across the face of the tilted disc.  The deeper in the primary's potential well the impact point on the face of the tilted disc, the brighter the instantaneous luminosity.  Simulations with mass ratios spanning the range $0.01\le q\le0.75$ and non-zero accretion rates through L1 produce a negative superhump signal in the light curve. The signals are antiphased for observers on opposite sides of the orbital plane.

The most surprising finding of this project was the discovery of the existence of a negative superhump signal in simulations for which the accretion stream had been shut off completely.  This previously unknown and unsuspected effect is only present in simulations with $q\ga0.30$. The signal has a fractional amplitude that peaks at $\sim$1\% for $q\approx0.35$, and falls gradually with increasing $q$.  We propose that the source of this signal is dissipation associated with forced driving of the $3:1$ vertical resonance identified by \citet{lubow92}.  The driving is strongest for lower mass ratios as these have most of the mass (particles) in the region of the 3:1 resonance.  The driving gets weaker with increasing $q$ as the driving frequency pulls away from the response frequency.

We give empirical fits to our simulation results which should be useful to observers working to identify negative superhumps in nature.  While most published negative superhump period deficits are in reasonable agreement with our numerical simulation results, a handful are in serious disagreement.  Because the physics of retrograde disc precession as it applies to these systems is likely to yield intrinsically small scatter about a well-defined trend, we suggest that it may be useful to observe the outlier systems with global campaigns in an effort to reduce the observational scatter.  If, however, new observations unambiguously confirm the large negative superhump period deficits reported for these systems, then perhaps the physics of photometric signal generation from tilted accretion discs in cataclysmic variable systems is even more interesting and complex than we currently know.  

\section*{Acknowledgments}

The lead author was on sabbatical from the Florida Institute of Technology (FIT) during the period of this work, and thanks Radboud University and The Netherlands Organisation for Scientific Research (NWO) for support through Visitor's Grant 040.11.046.  This was supported in part by the National Science Foundation through grants AST-0205902 and AST-0552798.
We thank Marcus Hohlmann and Patrick Ford from the FIT High Energy Physics group and the Domestic Nuclear Detection Office in the Dept. of Homeland Security for making computing resources on a Linux cluster available for this work.  Thanks to Michele Montgomery for writing a short Fortran code for rotating the particle positions and velocities that was used in this study.
This research has made use of NASA's Astrophysics Data System Bibliographic Services.

\label{lastpage}

\end{document}